\documentclass[a4paper,twocolumn,pra,showpacs,showkeywords,aps]{revtex4-1}
\usepackage{epsfig,amsmath,amsfonts,amssymb}
\usepackage{amsmath,amsfonts,amssymb}
\usepackage{graphicx}
\usepackage{mathbbol}
\newcommand{\beq}{\begin{equation}}
\newcommand{\eeq}{\end{equation}}

\newcommand{\beqa}{\begin{eqnarray}}
\newcommand{\eeqa}{\end{eqnarray}}
\newcommand{\beqas}{\begin{eqnarray*}}
\newcommand{\eeqas}{\end{eqnarray*}}
\def\ra{\rangle}
\def\la{\langle}

\begin{document}
\title{Optimal trajectories for efficient atomic transport without final excitation}
\author{Xi Chen$^{1,2}$}

\author{E. Torrontegui$^{1,3}$}

\author{Dionisis Stefanatos$^{4}$}

\author{Jr-Shin Li$^{4}$}

\author{J. G. Muga$^{1,3}$}

\affiliation{$^{1}$Departamento de Qu\'{\i}mica-F\'{\i}sica,
UPV-EHU, Apdo 644, 48080 Bilbao, Spain}

\affiliation{$^{2}$Department of Physics, Shanghai University,
200444 Shanghai, P. R. China}

\affiliation{$^{3}$Max Planck Institute for the Physics of Complex Systems,
N\"othnitzer Str. 38, 01187 Dresden, Germany}

\affiliation{$^{4}$Department of Electrical and System Engineering, Washington University, St. Louis, Missouri 63130, USA}

\begin{abstract}
We design optimal harmonic-trap trajectories to transport cold atoms without final excitation,
combining an inverse engineering techniqe based on Lewis-Riesenfeld invariants with optimal control theory. Since actual traps are not really harmonic,
we keep the relative displacement between the center of mass and the trap center bounded.
Under this constraint, optimal protocols are found according to different physical criteria.
The minimum time solution has a ``bang-bang" form, and the minimum displacement
solution is of ``bang-off-bang" form. The optimal trajectories for minimizing the
transient energy are also discussed.

\end{abstract}
\maketitle
\section{Introduction}
Efficient transport of ultracold atoms and ions by moving the confining trap is an important goal in atomic physics \cite{Leibfried2002,ions,HanschPRL2001,HanschNature2001,Ketterle2002,Denschlag,Lahaye,David,Calarco2009,Nakamura2010,Shan,James,transport,transport2},
with applications to basic science, metrology, and quantum information processing.
A sufficiently slow, adiabatic motion is a simple way to
transport the atoms without excitations or losses \cite{HanschNature2001,Ketterle2002,Denschlag}. However, the long time required may become impractical, e.g. if a fast quantum-information operation is required \cite{Leibfried2002,James};  or counterproductive, because of the accumulation of perturbations.
Motivated by this,
several theoretical and experimental investigations
have been devoted to making atomic transport fast, and simultaneously
faithful to the ideal result of adiabatic transport \cite{David,Calarco2009,Nakamura2010,transport,transport2,Shan}.

These works on transport share concepts and techniques with
other operations in which a ``shortcut to adiabaticity'' is to be found,
e.g. in expansion or compressions \cite{Salamon09,bec,Chen,Nakamura2010,JPB10,energy,Nice,Nice2,Wu,Adol},
rotations \cite{Nice3}, and internal state population transfer  \cite{Rice03,Rice05,Rice08,Berry09,Chen10b,Chen11,nonHermitian}.
Several approaches have been proposed, including
counter-diabatic \cite{Rice03,Rice05,Rice08} or, equivalently, transitionless driving algorithms \cite{Berry09,Chen10b,Chen11}, optimal control theory \cite{Salamon09}, ``fast-forward'' scaling  \cite{Nakamura2010}, and
inverse engineering based on Lewis-Riesenfeld invariants
\cite{LR,bec,Chen,JPB10,energy,Nice,Nice2,transport,Chen11,transport2}.

In essence, the invariant-based inverse engineering method relies on designing the Hamiltonian evolution so that the eigenvectors of  corresponding invariants of motion become at initial and final times equal to the instantaneous eigenvectors of the Hamiltonian. This method provides
in fact families of paths \cite{Chen11} which satisfy the initial and final boundary conditions, and thus guarantee the fast transitionless evolution, ideally in an arbitrarily short time.
Given this freedom, it is natural to combine the invariant-based inverse method and optimal control theory to optimize the trajectory according to different physical criteria or operational constraints. For example, the time-dependent frequency of a harmonic trap expansion can be optimized with respect to time or to transient excitation energy, with a restriction of the allowed
transient frequencies \cite{stef,stefanatosSICON,Li-energy}.

In this paper, we apply the invariant-based method complemented by optimal control theory to find optimal trajectories for fast atomic transport on harmonic traps without final vibrational excitation. Since actual traps are not really harmonic,
we keep, as an imposed constraint, the relative displacement between the center of mass and the trap center bounded.
We then optimize the trajectories according to different physical criteria: time minimization, (time averaged) displacement minimization, and (time averaged) transient energy minimization.
\section{Invariant-based inverse engineering method}
\label{sec2}
We consider here the harmonic transport described by the time-dependent Hamiltonian
\beq
\label{Hamiltonian}
H (t)= \frac{\hat{p}^2}{2 m} + \frac{1}{2}m \omega^2_0 [\hat{q}-q_0(t)]^2,
\eeq
where $\hat{q}$ and $\hat{p}$ are the position and momentum operators, $\omega_0$ is the constant harmonic frequency of the potential,
and $q_0(t)$ is the position of the center of the harmonic trap.
The corresponding
quadratic-in-momentum Lewis-Riesenfeld invariant \cite{LR} has the form \cite{LL,DL} (up to an arbitrary multiplicative constant)
\beq
\label{inva}
I (t) = \frac{1}{2m}(\hat{p}-m\dot{q}_c)^2 +\frac{1}{2}m \omega^2_0 [ \hat{q}-q_c (t)]^2,
\eeq
where the functions $q_c (t)$ must satisfy the auxiliary equation
\beq
\label{classical}
\ddot{q}_c+\omega_0^2(q_c-q_0)=0,
\eeq
to guarantee the invariant condition
\beq
\frac{d I(t)}{d t} \equiv \frac{\partial I(t)}{ \partial t} +\frac{1}{i \hbar} [I(t), H(t)] =0.
\eeq
Eq. (\ref{classical}) is simply Newton's equation for a classical particle
in the moving harmonic potential.

An arbitrary solution of the time-dependent Schr\"odinger equation
$i \hbar \partial_t\Psi (q,t) = H(t) \Psi (q,t)$,
may be written in terms of ``transport modes'' $e^{i\alpha_n} \psi_n(q,t)$,
\beqa
\Psi(q,t) &=& \sum_n c_n e^{i\alpha_n} \psi_n(q,t),
\eeqa
where $n=0,1,...$, $c_n$ are time-independent coefficients, $\psi_n(q,t)$ are the orthonormal eigenvectors of the invariant $I(t)$ satisfying $I(t)\psi_n(q,t)= \lambda_n\psi_n(q,t)$,
with real time-independent $\lambda_n$, and the Lewis-Riesenfeld phase is defined as
\beq
\label{LRphase}
\alpha_n (t) = \frac{1}{\hbar} \int_0^t \Big\langle \psi_n (t') \Big|
i \hbar \frac{\partial }{ \partial t'} - H(t') \Big| \psi_n (t')  \Big\rangle d t'.
\eeq
For the harmonic trap considered here \cite{DL},
%
\beqa
\label{psin}
\psi_n(q,t) &=& \frac{1}{(2^n n!)^{1/2}} \left( \frac{m \omega_0}{\pi \hbar}\right)^{1/4}  \exp{\left[- \frac{m \omega_0}{2\hbar} (q-q_c)^2\right]} \nonumber \\
 &\times& \exp{\left(i \frac{ m\dot{q}_c q}{\hbar}\right)}  H_n \left[\left(\frac{m \omega_0}{\hbar}\right)^{1/2}(q-q_c)  \right],
\eeqa
i.e., $q_c$ is the center of mass of the transport modes.
Substituting Eq. (\ref{psin}) into Eq. (\ref{LRphase}),
\beq
\label{alpha}
\alpha_n = -\frac{1}{\hbar}\int_0^t  {\rm d} t' \left(\lambda_n+\frac{m\dot{q}^2_c}{2} \right),
\eeq
where $\lambda_n=E_n=(n+1/2)\hbar \omega_0$.
The instantaneous average energy for a transport mode can be obtained from Eqs. (\ref{Hamiltonian}) and (\ref{psin}),
\beqa
\langle \psi_n(t)|H(t)|\psi_n(t)\rangle = {\hbar\omega_0}\left(n+1/2\right) + E_{c} + E_{p},
\eeqa
where the first, ``internal'' contribution remains constant for each $n$,
$E_{c} =m \dot{q}_c^2/2$, and $E_p= \frac{1}{2} m\omega_0^2(q_c-q_0)^2$ has the form of a potential energy for a classical particle.
The instantaneous average potential energy can be written as
\beq\label{potener}
\la V(t)\ra=\frac{\hbar\omega_0}{2}\left(n+1/2\right)+E_p.
\eeq

Suppose that the harmonic trap is displaced from $q_0 (0) =0$ to $q_0 (t_f) =d$ in a time $t_f$. The trajectory $q_0 (t)$ of the trap can be inverse engineered by designing first an appropriate classical
trajectory $q_c (t)$. To avoid vibrational excitation at the final time
we impose the conditions
\beqa
\label{con0}
q_c(0)=0;\; \dot{q}_c(0)=0;\; \ddot{q}_c(0)=0,
\\
\label{contf}
q_c(t_f)=d;\; \dot{q}_c(t_f)=0;\; \ddot{q}_c(t_f)=0,
\eeqa
which, along with Eq. (\ref{classical}), imply also
\beq
q_0(0)=0;\;q_0(t_f)=d.
\label{q0bc}
\eeq
The above boundary conditions guarantee the commutativity  of $I(t)$ and $H(t)$ at $t=0$ and $t=t_f$, that is,
the transport modes coincide with the eigenvectors of the instantaneous Hamiltonian at $t=0$ and $t=t_f$.
As discussed later in more detail the boundary conditions on the second derivatives, and consequently the conditions
for $q_0$ in Eq. (\ref{q0bc}) are special, in the sense that we shall allow for discontinuities in the acceleration $\ddot{q}_c$
at the edge times (in fact also elsewhere). Physically this means that the trap is ideally allowed to
be displaced suddenly a finite distance, inducing a sudden finite jump of the acceleration, whereas the velocity $\dot{q}_c$ and the trajectory $q_c$ remain always continuous.
$q_c (t)$ can be interpolated by a simple polynomial ansatz that satisfies these boundary conditions.
Once $q_c (t)$ is fixed, we get the trap trajectory $q_0 (t)$ from Eq. (\ref{classical}).
In principle there is no lower bound for $t_f$ \cite{transport}.
However, there are always some limits in the laboratory related, for instance,
to spatial or energy constraints.
%
%
%
%
%
%
\section{Optimal control problem with constrained relative displacement}
We begin with the equation of motion, Eq. (\ref{classical}), for the classical particle
in the harmonic trap, and set, for compactness and to follow the usual conventions in optimal control theory, a new notation,
\beqa
x_1 = q_c,~ x_2 = \dot{q}_c, ~u(t) = q_c -q_0,
\eeqa
where $x_1, x_2$ are the components of a ``state vector'' $\bf{x}$, and the relative displacement between the trap and the center of mass $u (t)$ is considered
as the (scalar) control function. The physical motivation behind this control is that actual traps are not really harmonic,
so the relative displacement should be kept bounded.
Eq. (\ref{classical}) becomes
\beqa
\label{system-1}
\dot{x}_1  &=&  x_2,
\\
\label{system-2}
\dot{x}_2 &=& - \omega_0^2 u.
\eeqa
%
%
The optimal control problem is to
find $|u (t)| \leq \delta$ for some fixed bound $\delta$,
with $u(0)=0$ and $u(t_f) =0$  such that the system starts
at $\{x_1(0)=0, x_2 (0)= 0\}$,
ends up at $\{x_1(t_f)=d, x_2 (t_f)= 0\}$, and minimizes a cost function $J$.

The boundary conditions for $x_1$ and $x_2$ can be equivalently considered as those for $q_c$ and $\dot{q}_c$.
The boundary conditions for $u(t)$
are equivalent to those for $q_0$ and, through Eq. (\ref{classical}), equivalent to those for $\ddot{q}_c$,
so there are totally six boundary conditions, as in Eqs. (\ref{con0}) and (\ref{contf}).
A natural way to understand the boundary conditions on $u(t)$
is to consider that $u(t)=0$ for $t\leq 0$ and $t\geq t_f$, so the center of mass and the trap center coincide
before and after the transport.
We will consider cost functions that are not affected by the isolated values $u(0)$ and $u(t_f)$,
for example minimizing the transport time or the energy, and solve the control problem in the interval $(0,t_f)$.
In order to match the boundary conditions at the initial and final times, the optimal control obtained may be complemented by appropriate jumps at these points which do not affect the cost.
We use Pontryagin's maximum principle, which provides necessary conditions for optimality \cite{LSP}.
Generally, to minimize the cost function
\beq
J (u)= \int^{t_f}_0 g [\textbf{x}(t), u] dt,
\eeq
the maximum principle states that for the dynamical system
$
\dot{\textbf{x}} = \textbf{f} [\textbf{x}(t),u],
$
the coordinates of the extremal vector $\textbf{x} (t)$ and of the corresponding adjoint state $\textbf{p} (t)$ formed by Lagrange multipliers, $p_1$, $p_2$, fulfill the Hamilton's equations
for a control Hamiltonian $H_c$,
\beqa
\label{H-1}
\dot{\textbf{x}} = \frac{\partial H_c}{\partial \textbf{p}},
\\
\label{H-2}
\dot{\textbf{p}} = - \frac{\partial H_c}{\partial \textbf{x}},
\eeqa
where $H_c$ is defined as
\beq
H_c [\textbf{p}(t),\textbf{x}(t),u] = p_0 g [\textbf{x}(t), u]+ \textbf{p}^{T}\cdot \textbf{f} [\textbf{x}(t),u].
\eeq
The superscript ``$T$" used here denotes the transpose of a vector,  and $p_0< 0$ can be chosen for convenience since it amounts to multiply the cost function by a constant. The (augmented) vector with components $(p_0, p_1, p_2)$ is nonzero and continuous.
For almost all $0 \leq t \leq t_f$ the function $H_c [\textbf{p}(t),\textbf{x}(t),u]$ attains its maximum at $u=u(t)$, and
$H_c [\textbf{p}(t),\textbf{x}(t),u(t)] =c$, where $c$ is constant.
%
%
%
%
\subsection{Time minimization}
We discuss now the time-minimization optimal control problem with a constrained relative displacement, that is,
$|u(t)| = |q_c - q_0| \leq \delta$, which means $E_p \leq \frac{1}{2} m \omega^2_0 \delta^2$. To find the minimal time $t_f$ we define the cost function
\beq
J_{T} = \int^{t_f}_0 dt = t_f.
\eeq
The control Hamiltonian $H_c [\textbf{p}(t),\textbf{x}(t),u]$ is
\beq
\label{Htime}
H_c (p_1, p_2, x_1, x_2, u)= p_0 + p_1 x_2 - p_2 \omega^2_0 u.
\eeq
%
With the control Hamiltonian,
Eq. (\ref{H-2}) gives the following costate equations,
\beqa
\label{costate time-1}
\dot{p}_1 &=& 0,
\\
\label{costate time-2}
\dot{p}_2 &=& -p_1.
\eeqa
They are solved easily as
$p_1 = c_1$ and $p_2 = -c_1 t +c_2$ with constants $c_1$ and $c_2$.
According to the Pontryagin's maximum principle, the time-optimal control $u(t)$ maximizes the control Hamiltonian in Eq. (\ref{Htime}).

\begin{figure}[t]
\begin{center}
\scalebox{0.6}[0.6]{\includegraphics{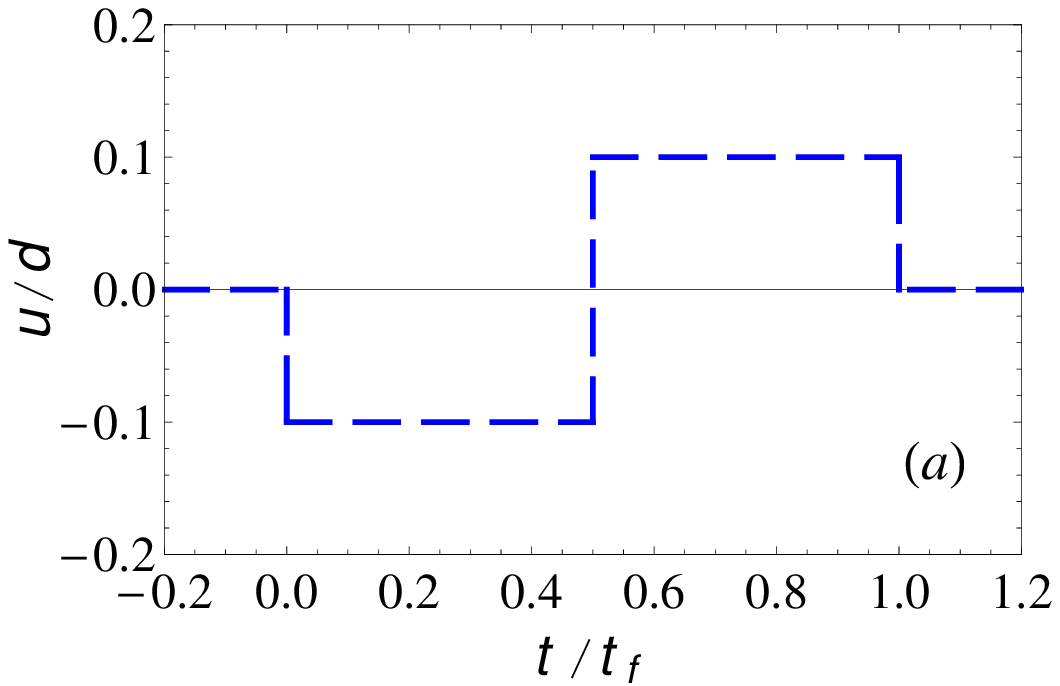}}
\scalebox{0.6}[0.6]{\includegraphics{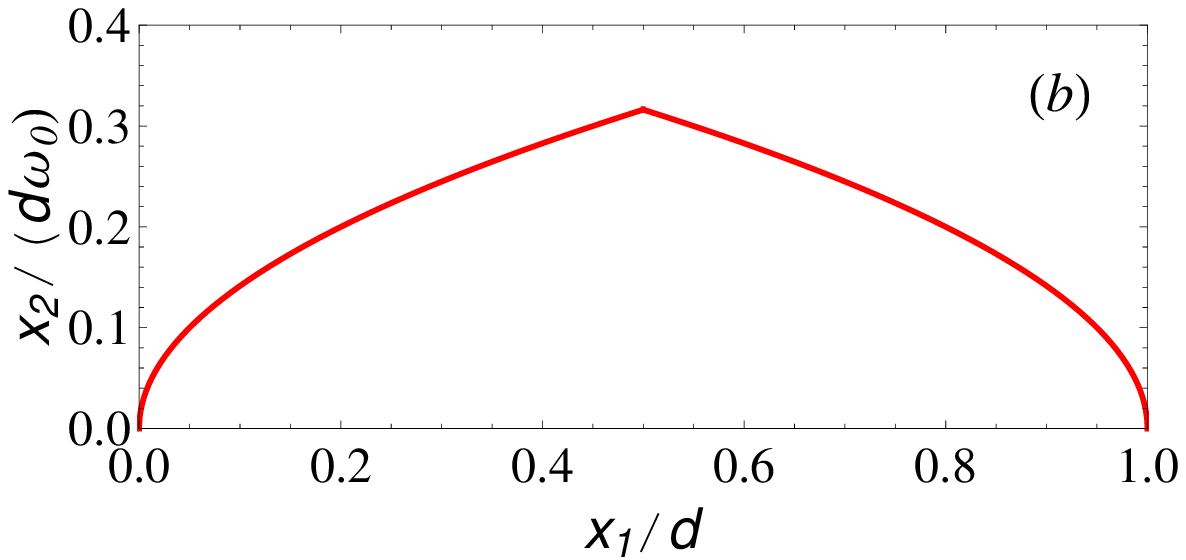}}
\caption{(Color online) (a) The control function (dashed blue line) for time-optimal problem, and (b) the corresponding trajectory (solid red line) for $\omega_0 = 2 \pi \times 50$ Hz, $d=1.6$ mm, $\delta=0.16$ mm and $t_f = 20$ ms given by Eq. (\ref{tf-bangbang}). }
\label{control-time}
\end{center}
\end{figure}

Since the control Hamiltonian is a linear function of the control
function $u(t)$, the optimal control that maximizes $H_c$ is determined by the sign of $p_2$, when $u (t)$ is bounded, $|u(t)| \leq \delta$. When $p_2 \neq 0$, the optimal control
in the duration $t_f$ is given by
\beqa
u (t) = \left\{\begin{array}{ll} - \delta, & p_2 > 0
\\
\delta, & p_2 <0
\end{array}\right..
\eeqa
If $p_2=0$ for some time interval, then $p_1=0$ from Eq. (\ref{costate time-2}), and $p_0=0$ from Eq. (\ref{Htime}), since $H_c=0$ for the time optimal problem \cite{LSP}, in contrast with the maximum principle that requires $(p_0,p_1,p_2)\neq 0$. Thus   $p_2$ can be zero only at isolated points, the switching times.
The solutions of the costate functions in Eqs. (\ref{costate time-1}) and (\ref{costate time-2})
imply that the function of $p_2$ depends linearly on time $t$, so that the sign of $p_2$ cannot change more than once.
Since the final point is $(x_1,x_2)=(d,0), d>0$, the appropriate control sequence is of ``bang-bang" (piecewise constant) type,
\beqa
\label{control function-time}
u (t) = \left\{\begin{array}{ll}
0, & t \leq 0
\\
- \delta, & 0<t <t_1
\\
\delta, & t_1< t <t_f
\\
0, & t \geq t_f
\end{array}\right.,
\eeqa
with only one intermediate switching time at $t_1$, as shown in Fig. \ref{control-time} (a).
The saturation of the control is typical of time minimization problems.

\begin{figure}[t]
\begin{center}
\includegraphics[width=0.8\linewidth]{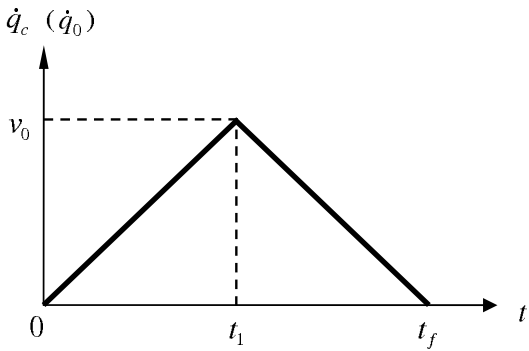}
\caption{Velocity profiles of harmonic trap and particles for time minimization,
where the switching time $t_1 =t_f/2$.}
\label{profile-bangbang}
\end{center}
\end{figure}

Substituting $u(t)$ into the classical Eq. (\ref{classical}),
and using the boundary conditions in Eqs. (\ref{con0}) and (\ref{contf}), we find
the optimal classical trajectory
\beqas
q_c (t)=\left\{
\begin{array}{ll}
0,& t \leq 0
\\
{\omega_{0}^{2}\delta t^2 }/{2},& 0<t<t_1
\\
d-{\omega_{0}^2}\delta(t-t_f)^2 /2,& t_1<t<t_f
\\
d,& t\geq t_f
\end{array}
\right.,
\eeqas
and the corresponding trajectory for the harmonic trap is
\beqas
q_0 (t)=\left\{
\begin{array}{ll}
0,& t \leq 0
\\
({1+\omega_{0}^{2}t^2}/{2})\delta,& 0<t<t_1
\\
d-\big[{\omega_{0}^2}(t-t_f)^2/2+1\big]\delta,& t_1<t<t_f
\\
d,& t \geq t_f
\end{array}
\right..
\eeqas
Fig. \ref{control-time} (b) illustrates the time-optimal trajectory with one switching time. Solving the system of Eqs. (\ref{system-1}) and (\ref{system-2}), one can find
the switching time $t_1$ and final time $t_f$,
\beqa
t_1 &=& \frac{t_f}{2},
\\
\label{tf-bangbang}
t_f &=& \frac{2}{\omega_0} \sqrt{\frac{d}{\delta}},
\eeqa
by imposing continuity on $x_1$ and $x_2$. For the ``bang-bang" control the motion of the trap has
discontinuities, while we impose continuity for the trajectory of the particle. As illustrated by Fig. \ref{profile-bangbang},
the velocities of particle and trap become
equal,
\beqa
\dot{q}_c = \dot{q}_0 = \left\{
\begin{array}{ll}
{\omega_{0}^{2} \delta t },& 0<t<t_1
\\
-{\omega_{0}^2}\delta(t-t_f) ,& t_1<t<t_f
\end{array}
\right.,
\eeqa
since $u(t)$ is piecewise constant during the ``bang-bang" control.
The maximum velocity occurs at $t=t_f/2$,
\beq
v_0 =\omega_{0}^{2}  \delta t_f/2 = \omega_{0} \sqrt{d \delta},
\eeq
which is restricted by the imposed bound $|u(t)| \leq \delta$. In addition,
the instantaneous potential energy $\la V \ra$ is constant, and
\beq
E_p = \frac{1}{2} m\omega_0^2 \delta^2 = \frac{8 m d^2}{\omega_0^2 t^4_f}.
\eeq
If we loosen the bound by increasing $\delta$,
the maximum velocity and the instantaneous potential energy increase, and
the final time may be shortened.

%
%
%
%
%
\subsection{Displacement minimization}
\begin{figure}[t]
\begin{center}
\scalebox{0.6}[0.6]{\includegraphics{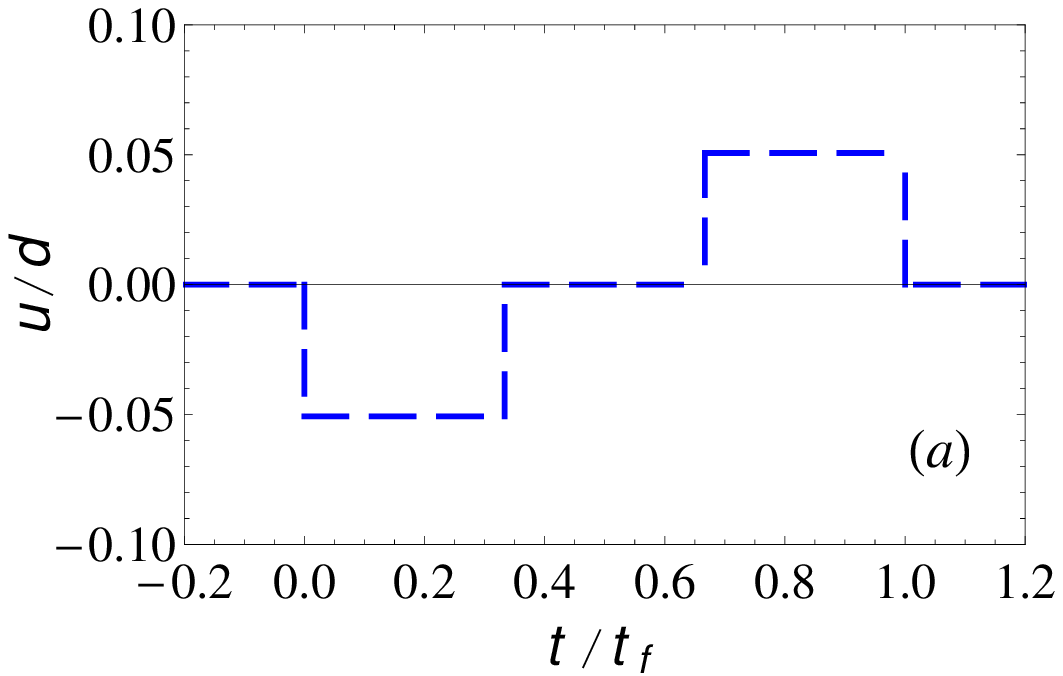}}
\scalebox{0.6}[0.6]{\includegraphics{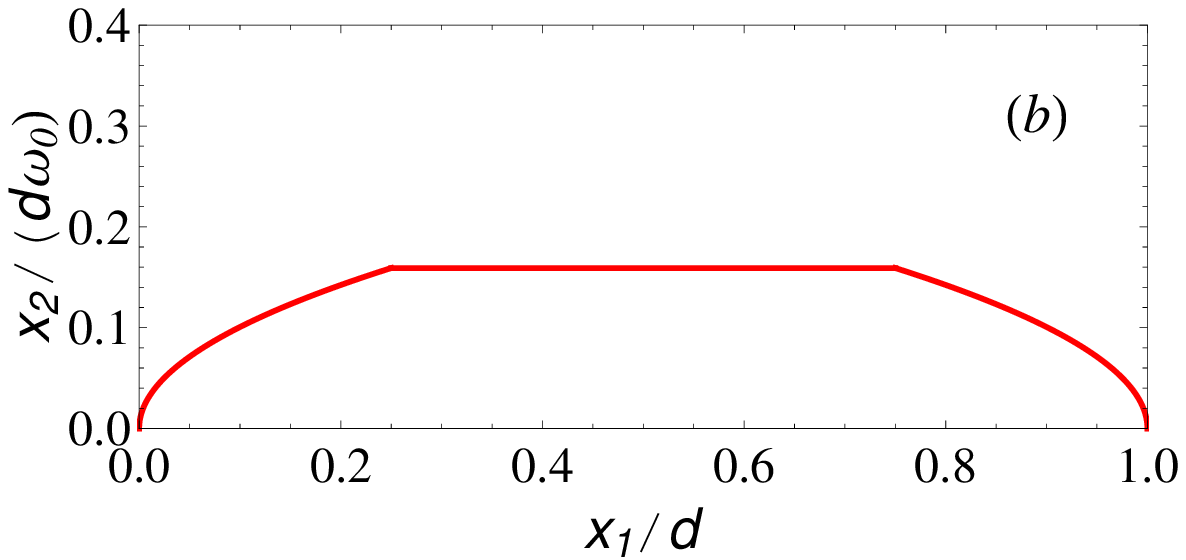}}
\caption{(Color online) (a) The control function (dashed blue line) for displacement-optimal problem, and (b) the corresponding trajectory (solid red line) for $t_f=30$ ms,
$\delta =9 d/2\omega^2_0 t^2_f$, and other parameters are
the same as Fig. \ref{control-time}.}
\label{control-displacement}
\end{center}
\end{figure}

In this subsection, we minimize the integral, or time-average of the
relative displacement, which is equivalent to a minimal control-effort problem.
To this end, the cost function can be defined as
\beq
\label{costfuntion-D}
J_{D} = \int^{t_f}_0 |u(t)| dt =  \int^{t_f}_0 |q_c - q_0| dt,
\eeq
and the control Hamiltonian is
\beq
\label{HD}
H_c (p_1, p_2, x_1, x_2, u)= p_0|u|+p_1x_2-p_2\omega_0^2 u,
\eeq
which leads to the same costate equations for $p_1 (t)$  and $p_2 (t)$ in Eqs. (\ref{costate time-1}) and (\ref{costate time-2}).
We use for convenience the normalization $p_0=-\omega_0^2$.
Disregarding $u$-independent terms in $H_c$, the function of $u(t)$ that
we have to maximize is
\beqa
-\omega_0^2 (|u| + p_2  u)  = \left\{\begin{array}{ll}
- \omega_0^2 (1  + p_2)u, & u \geq  0
\\
\omega_0^2 (1 - p_2)u, & u \leq 0
\end{array}\right..
\eeqa
According to Pontryagin's maximum principle, when $u(t)$ is bounded, $|u(t)| \leq \delta$, the control function is
\beqa
u (t) = \left\{\begin{array}{lll}
- \delta, & p_2 > 1
\\
0, & -1< p_2 < 1
\\
\delta, & p_2 < -1
\end{array}\right.,
\eeqa
which maximizes the control Hamiltonian in Eq. (\ref{HD}).
Notice that whereas in the minimum-time problem discussed above,
the optimal control is ``bang-bang", the minimal-displacement control can be described as ``bang-off-bang", if we assume
no singular intervals here. Owing to the properties of costate equations, the ``bang-off-bang" trajectory with two switching times
$t_1$ and $t_1+t_2$ can be described by, see Fig. \ref{control-displacement} (a),
\beqa
u (t) = \left\{\begin{array}{lll}
0, & t \leq 0
\\
- \delta, & 0<t <t_1
\\
0, & t_1<t<t_1+ t_2
\\
\delta, & t_1 + t_2 < t <t_f
\\
0, & t \geq t_f
\end{array}\right..
\eeqa
%

\begin{figure}[t]
\begin{center}
\includegraphics[width=0.8\linewidth]{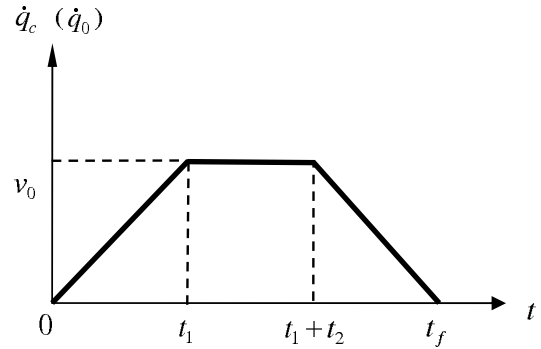}
\caption{Velocity profiles of harmonic trap and particles for displacement-minimization with two switching times $t_1$ and $t_1+ t_2$.}
\label{profile-bang off bang}
\end{center}
\end{figure}

Substituting the control function $u(t)$ into the classical Eq. (\ref{classical}), using the boundary conditions in Eqs. (\ref{con0}) and (\ref{contf}), and
imposing the continuity of $q_c$ at the two switching times, the optimal trajectory for
the center of mass, as shown in Fig. \ref{control-displacement} (b), is finally given by
\beqas
q_c (t)= \left\{
\begin{array}{lll}
0,& t \leq 0
\\
{\omega_{0}^{2} \delta t^2 }/{2},& 0<t<t_1
\\
v_0 t - v_0^2/(2 \omega^2_0 \delta), & t_1<t<t_1+t_2
\\
d-{\omega_{0}^2}\delta(t-t_f)^2 /2,& t_1 +t_2 < t < t_f
\\
d, & t \geq t_f
\end{array}
\right.,
\eeqas
which results in the following optimal trap trajectory,
\beqas
q_0 (t)= \left\{
\begin{array}{lll}
0,& t \leq 0
\\
(1+{\omega_{0}^{2} t^2}/{2}) \delta,& 0<t<t_1
\\
v_0 t - v_0^2/(2 \omega^2_0 \delta), & t_1<t<t_1+t_2
\\
d-\big[{\omega_{0}^2}(t-t_f)^2 /2+1\big] \delta,& t_1 +t_2 < t < t_f
\\
d, & t \geq t_f
\end{array}
\right..
\eeqas
As in the ``bang-bang" time-minimization, the velocities of particle and trap
are equal, see Fig. \ref{profile-bang off bang},
\beqa
\dot{q}_c = \dot{q}_0 = \left\{
\begin{array}{lll}
{\omega_{0}^{2} \delta t},& 0<t<t_1
\\
v_0, & t_1<t<t_1+t_2
\\
-{\omega_{0}^2}\delta(t-t_f) ,& t_1+t_2<t<t_f
\end{array}
\right.,
\eeqa
where $v_0$ is the maximum velocity of trap motion in the trajectory, which will be determined later.
With the boundary conditions for $x_1$ and $x_2$ at $t=t_1$ and $t=t_1+t_2$,
the switching times can be calculated as
\beqa
\label{switching time-1}
t_1 &=& \frac{v_0}{\omega^2_0 \delta},
\\
\label{switching time-2}
t_2 &=& \frac{d}{v_0}- \frac{v_0}{\omega^2_0 \delta}.
\eeqa
As a consequence, the final time is
\beq
\label{time-d}
t_f= 2t_1 +t_2 = \frac{d}{v_0} + \frac{v_0}{\omega_0^2  \delta} \geq \frac{2}{\omega_0}\sqrt{\frac{d}{\delta}}.
\eeq
%
Since the final time $t_f$ is fixed, there are three possible cases:
({\it i}) when $t_f > (2/\omega_0)\sqrt{d/\delta}$, the maximal velocity $v_0$ can be solved
from Eq. (\ref{time-d}) as
\beq
\label{velocity-d}
v^{\pm}_0 = \frac{\omega^2_0 \delta t_f}{2} \left( 1 \pm  \sqrt{1 -  \frac{4d}{ \omega^2_0  t^2_f \delta}} \right),
\eeq
where $v^{+}_0$ should be ignored, because it leads to $2t_1>t_f$.
({\it ii}) If $t_f = (2/\omega_0)\sqrt{d/\delta}$, the maximum velocity
is $v_0= \omega_0 \sqrt{d\delta}$, thus $t_1= t_f/2$ and $t_2=0$. The trajectory in this case
is reduced to that of the time-optimal control problem.
({\it iii}) When the time $t_f$ is less than $(2/\omega_0)\sqrt{d/\delta}$,
there is no real solution to $v_0$ and no solution to
displacement minimization.

Interestingly, the ``bang-off-bang" trajectory obtained from displacement minimization
may be related to the trajectory used for atomic transport in \cite{HanschPRL2001,HanschNature2001}, where the
shift velocity $\dot{q}_0$ was increased linearly during a quarter of the spatial transported distance $d/4$, then kept constant for $d/2$,
and during the last quarter finally ramped back to zero. To understand this in the context of optimal control theory for fixed
$t_f$ and $d$, we note that for the choice
\beq\label{cH}
\delta = \frac{9d}{2 \omega^2_0 t^2_f},
\eeq
the maximum velocity  $v_0$ is, according to Eq. (\ref{velocity-d}),
\beq
v_0 = \frac{3 d}{2 t_f},
\eeq
and the switching times in Eqs. (\ref{switching time-1}) and (\ref{switching time-2}) become
\beqa
\label{switching time}
t_1 = \frac{1}{3} t_f = \frac{d}{2 v_0},~ t_2  = \frac{1}{3} t_f = \frac{d}{2 v_0}.
\eeqa
The positions of the classical particle at the two switching times are
\beq
q_c (t_1) = d/4, ~~ q_c(t_1+t_2)=3d/4.
\eeq
Due to the discontinuity at $t=0$ and $t=t_f$, the motion of trap begins with $q_0(t=0^{+}) = \delta$ and
ends up with $q_0 (t=t^{-}_f) = d -\delta$, so that
\beq
q_0 (t^{-}_1) =d/4+ \delta, ~~ q_0 [(t_1+t_2)^{+}] = 3d/4 - \delta.
\eeq
In other words, the protocol followed in \cite{HanschPRL2001,HanschNature2001} minimizes the averaged displacement by imposing the bound in Eq. (\ref{cH}) to the displacement.

Returning now to the general case, the time-averaged potential energy for the optimal
trajectory is
\beq
\overline{E_p}= \frac{\int^{t_f}_0 E_p dt}{t_f} = \frac{m \omega^2_0 \delta^2 t_1}{t_f},
\eeq
where $t_1=v_0/ \omega^2_0 \delta$ is given by
\beq
t_1 = \frac{t_f}{2} \left( 1 - \sqrt{1 -  \frac{4d}{ \omega^2_0  t^2_f \delta}} \right).
\eeq
As a result,
\beq
\label{energy-d}
\overline{E_p}=  \frac{1}{2} m \omega^2_0 \delta^2 \left( 1 - \sqrt{1 -  \frac{4d}{ \omega^2_0 t^2_f \delta}} \right).
\eeq
For example, when $\delta = 9d/ 2 \omega^2_0 t^2_f$ and $t_1=t_f/3$ are chosen as discussed above, the time-averaged potential energy
is $\overline{E_p}= 27 m d^2/ 4 \omega^2_0 t^4_f$, which is less than the
(constant) potential energy
${E_p} = 8 m d^2/ \omega^2_0 t^4_f $ for the time-optimal control problem.
\subsection{Energy minimization}
%
%
%
%
%
The instantaneous potential energy
$\la V(t)\ra$ is given in Eq. (\ref{potener}).
To minimize the potential energy average for a given $n$ and fixed transport time $t_f$, the cost function can be defined as
\beq
\label{costfuntion-E}
J_{E} =  \int^{t_f}_0 E_p dt= \int^{t_f}_0 \frac{1}{2} m \omega^2_0 u^2 dt,
\eeq
and the control Hamiltonian is
\beq
H_c = - p_0 \frac{1}{2} m \omega^2_0 u^2 + p_1 x_2 - p_2 \omega^2_0 u,
\eeq
which gives two costate equations, Eqs. (\ref{costate time-1}) and (\ref{costate time-2}). The solutions are $p_1 = c_1$ and
$p_2 = -c_1 t + c_2$, with constat $c_1$ and $c_2$.
For the normalization $p_0=-1/m$ the function of $u(t)$ that we have to maximize is $-u^2/2-p_2u$.

\begin{figure}[t]
\begin{center}
\scalebox{0.6}[0.6]{\includegraphics{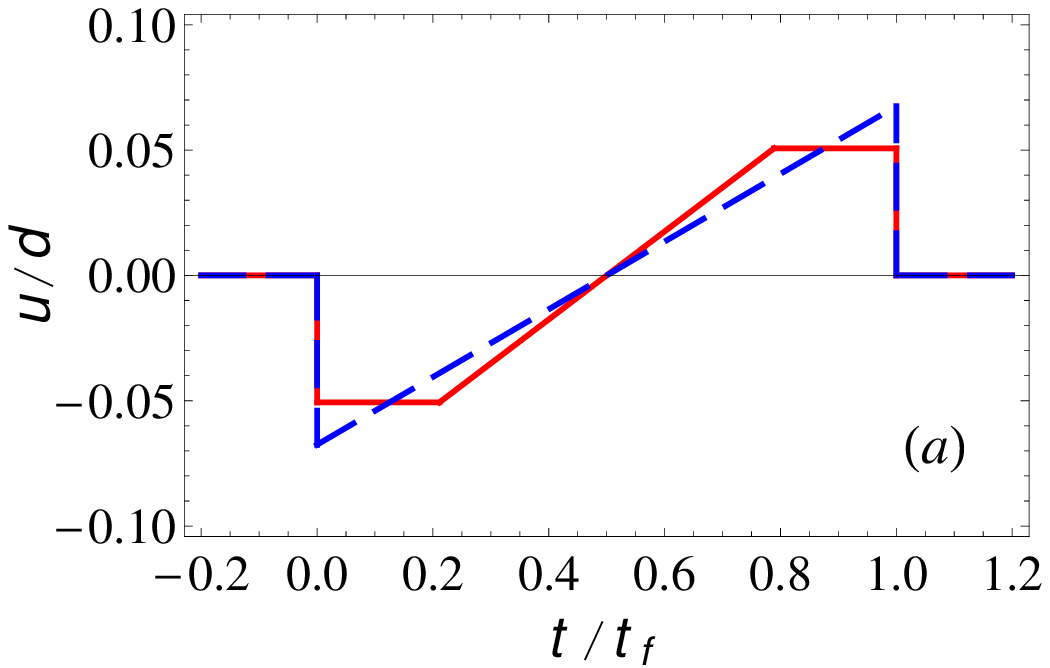}}
\scalebox{0.6}[0.6]{\includegraphics{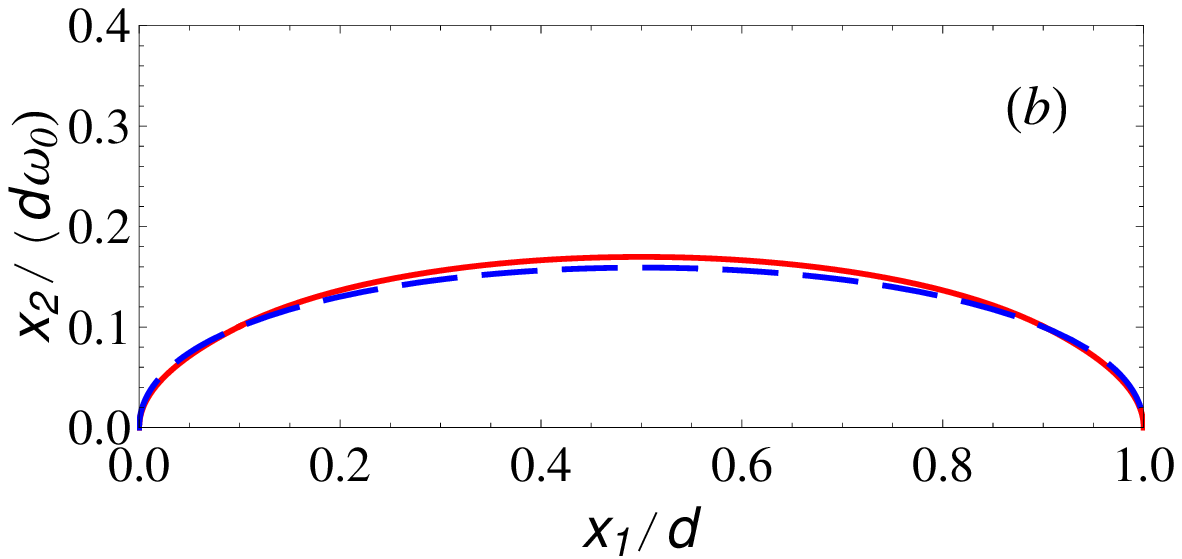}}
\caption{(Color online) (a) Control functions for the energy-optimal problem with bounded  (solid red line) and unbounded (dashed blue line) controls, and (b) corresponding trajectories in the cases of bounded (solid red line)  and unbounded  (dashed blue line) controls. The parameters are
the same as in Fig. \ref{control-displacement}.}
\label{control-energy}
\end{center}
\end{figure}

Here we start with  the  case  of  ``unbounded control", i.e.,
without imposing any constraints on the displacement, and we
shall show how this is related to the  physically interesting case where the control is bounded.
To maximize $-u^2/2-p_2u$, the control function is found to be 
\beqa
u (t) = -p_2,
\eeqa
%
%
and the classical Eq. (\ref{classical}),
$\ddot{q}_c = -\omega^2_0 u$, gives the optimal trajectory
\beqa
\label{quasi-trajectory}
q_c =
-\frac{1}{6} c_1 \omega^2_0 t^3 + \frac{1}{2} c_2 \omega^2_0 t^2 +c_3 t+ c_4.
\eeqa
%
%
Using the boundary conditions for $q_c$ and $\dot{q}_c$ in Eqs. (\ref{con0}) and (\ref{contf}),
we find $c_1=12d/\omega^2_0 t^3_f$, $c_2=6d/\omega^2_0 t^2_f$, $c_3=0$ and $c_4=0$.
Clearly, Eq. (\ref{quasi-trajectory}) does not satisfy the boundary conditions for $\ddot{q}_c$ in Eqs. (\ref{con0}) and (\ref{contf}).
To guarantee $u(t)=0$ at $t \leq 0$ and $t \geq t_f$ and match the boundary conditions, the control function $u(t)$ has to be complemented by the
appropriate jumps at these two edges.
Consequently, the control function for unbounded control, see Fig. \ref{control-energy} (a), is found to be
\beqa
\label{control function-unbounded}
u (t) = \left\{\begin{array}{lll}
0, & t \leq 0
\\
\frac{6d}{\omega_0^2t_f^2}\left(2\frac{t}{t_f}-1\right), & 0<t <t_f
\\
0, & t \geq t_f
\end{array}\right..
\eeqa
As shown in Fig. \ref{control-energy} (b), the optimal classical trajectory for unbounded control finally becomes
\beqa
\label{trajectory-unbounded}
q_c = \left\{\begin{array}{lll}
0, & t \leq 0
\\
\frac{d t^2}{t^2_f} \left(3 -2 \frac{t}{t_f}\right), & 0<t <t_f
\\
d, & t \geq t_f
\end{array}\right.,
\eeqa
where the trajectory $q_c$ in the interval $(0,t_f)$ is in agreement with the result obtained in \cite{transport}
using the Euler-Lagrange equation.
In this case, the time-averaged minimal potential energy is
\begin{equation}
\label{energy bound}
\overline{E_p}^{min}= \frac{\int^{t_f}_0 E_p dt}{t_f}=\frac{6md^2}{\omega_0^2t_f^4},
\end{equation}
which gives a lower bound for the time averaged potential energy of any other trajectories satisfying all the boundary conditions,
$
\overline{E_p} \geq 6md^2/ \omega_0^2t_f^4.
$
Note that, in spite of not having preimposed a bound for the displacement,
the optimal trajectory obeys $|u(t)|\leq\delta_0=6d/\omega_0^2t_f^2$.
For the bounded control, i.e., when $|u(t)|\leq\delta$
is imposed, if $\delta\geq\delta_0$ the unbounded solution is the optimal one.
(The value of $\delta_0$ can be obtained in the bounded control case
by requiring $t_1 \geq 0$, see Eq. (\ref{time-energy1}) below.)

When the bound, $|u(t)| \leq \delta$, is imposed,
the control function is
\beqa
u (t) = \left\{\begin{array}{lll}
- \delta, & ~~~~ p_2 >  \delta
\\
-p_2, & -\delta < p_2 < \delta
\\
\delta, &~~~ p_2 < - \delta
\end{array}\right.,
\eeqa
to achieve the maximum value of the control Hamiltonian $H_c$.
As before, the linear $p_2$ implies two switching times
$t_1$ and $t_1+t_2$. To make the control function continuous at
$t_1$ and $t_1+t_2$, it has the form shown in Fig. \ref{control-energy} (a),
\beqa
\label{control function-energy}
u (t) = \left\{\begin{array}{lll}
0, & t \leq 0
\\
- \delta, & 0<t <t_1
\\
c_1 (t -t_f/2), & t_1<t<t_1+ t_2
\\
\delta, & t_1 + t_2 < t <t_f
\\
0, & t \geq t_f
\end{array}\right.,
\eeqa
where, because of $t_f = 2 t_1 + t_2$ due to the symmetry, the two switching times $t_1$
and $t_2$ are given by
\beqa
t_2= 2 \delta /c_1,~~~ t_1= \frac{t_f-2 \delta/c_1}{2}.
\eeqa
%

\begin{figure}[t]
\begin{center}
\includegraphics[width=1.0\linewidth]{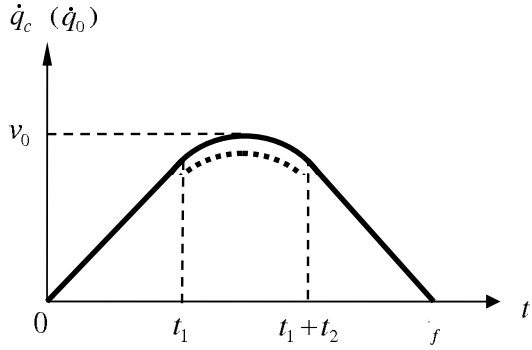}
\caption{Velocity profiles of harmonic trap (dashed line) and center of mass (solid line) for energy-minimization with two switching times $t_1$ and $t_1+ t_2$.}
\label{profile-energy}
\end{center}
\end{figure}

\begin{figure}[t]
\begin{center}
\includegraphics[width=0.8\linewidth]{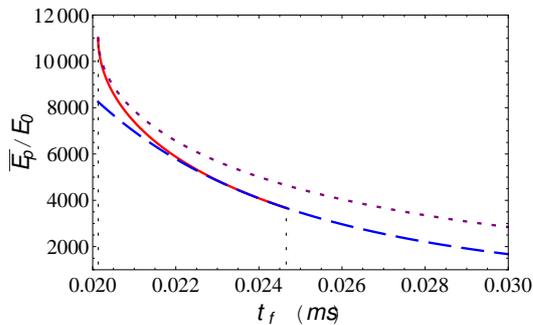}
\caption{Dependences of time-averaged energies $\overline{E_p}$ on $t_f$ with $\delta =0.1 d $, for energy-minimization with bounded control (red solid line), unbounded control, Eq. (\ref{energy bound}) (blue dashed line),
and displacement-minimization (purple dotted line).
Parameters are $d=1.6$ mm, $\omega_0 = 2 \pi \times 50$ Hz, $E_0 = 100 \times (\hbar \omega_0/2)$ and
mass of 
Rubidium 87.
}
\label{energy}
\end{center}
\end{figure}

Unlike the time-minimization and displacement-minimization problems,
the control function here is not piecewise constant,
so the velocities of the classical particle and the trap are not equal during the second segment from $t_1$ to $t_1+t_2$,
see Fig. \ref{profile-energy}. According to the control function in Eq. (\ref{control function-energy}), imposing the boundary conditions
for $x_2$ at $t=0$ and $t=t_f$, the velocity for
the center of mass is
\beqa
\dot{q}_c = \left\{
\begin{array}{lll}
{\omega_{0}^{2} \delta t},& 0<t<t_1
\\
-\frac{1}{2}\omega_{0}^{2} c_1 ( t-\frac{t_f}{2})^2 +v_0, & t_1<t<t_1+t_2
\\
-{\omega_{0}^2}\delta(t-t_f) ,& t_1+t_2<t<t_f
\end{array}
\right.,
\label{deriv}
\eeqa
and $\dot{q}_0 = \dot{q}_c - \dot{u}$ gives the velocity profile of the trap,
\beqa
\nonumber
\dot{q}_0 &=& \left\{
\begin{array}{lll}
{\omega_{0}^{2} \delta t},& 0<t<t_1
\\
-\frac{1}{2}\omega_{0}^{2} c_1 (t-\frac{t_f}{2})^2 +v_0-c_1, & t_1<t<t_1+t_2
\\
-{\omega_{0}^2}\delta(t-t_f) ,& t_1+t_2<t<t_f
\end{array}
\right.,
\\
\eeqa
where $v_0$ is the maximum velocity.
With $t_2= 2\delta/c_1$, and further imposing continuity of $x_2$ at
$t=t_1$ and $t=t_1+t_2$, we find
\beq
\label{Etime-1}
t_1 = \frac{v_0}{\omega^2_0 \delta} -\frac{\delta}{2 c_1},
\eeq
which finally leads to  $t_f= 2 t_1 + t_2$,
\beq
\label{Etf}
t_f = \frac{2 v_0}{\omega^2_0 \delta}+\frac{\delta}{c_1}.
\eeq
Solving Eqs. (\ref{Etime-1}) and (\ref{Etf}), the parameters $c_1$ and $v_0$ are given by
\beq
c_1= \frac{ 2 \delta}{t_f - 2t_1}, ~~ v_0 = \frac{1}{4} \omega^2_0 \delta (t_f +2 t_1).
\eeq
Thus,
$
c_2 = \delta t_f/(t_f -2 t_1)
$.
So far, $c_1$, $c_2$ and $v_0$ are all functions of $t_1$.
To determine $t_1$
we write down the optimal-energy classical trajectory
from Eq. (\ref{deriv}),
\beqas
q_c (t) &=& \left\{
\begin{array}{lll}
0,& t \leq 0
\\
{\frac{1}{2}\omega_{0}^{2} t^2 \delta}, & 0<t<t_1
\\
- \frac{1}{6}\omega^2_0 c_1 (t-\frac{t_f}{2})^3 +v_0 t+c_3, & t_1<t<t_1+t_2
\\
d-\frac{1}{2}{\omega_{0}^2}(t-t_f)^2 \delta,& t_1 +t_2 < t < t_f
\\
d, & t \geq t_f
\end{array}
\right..
\eeqas
By using the continuity of $x_1$ at $t=t_1$ and $t=t_1+t_2$, $c_3$ and $t_1$ can be solved as
\beqa
c_3 &=& \frac{1}{2} (d-v_0 t_f),
\\
\label{time-energy1}
t_1 &=& \frac{t_f}{2} \left( 1- \sqrt{3} \sqrt{1 - \frac{4 d}{\omega^2_0 t^2_f \delta}}\right),
\eeqa
where the other unphysical solution should be neglected.
Once $t_1$ is fixed, $c_j$ ($j=1,2,3$) are available, and $v_0$ is given by
\beq
\label{velocity}
v_0 = \frac{\omega^2_0 \delta t_f}{2} \left( 1 -\frac{\sqrt{3}}{2}  \sqrt{1 -  \frac{4d}{ \omega^2_0  t^2_f \delta}} \right),
\eeq
which is less than the maximum velocity for the displacement-optimal trajectory.
A trajectory with minimal energy and bounded control is depicted in Fig. \ref{control-energy} (b). It is seen from Eqs. (\ref{time-energy1}) and (\ref{velocity}) that
for a real $t_1$ and $v_0$, $t_f \geq (2/\omega_0)\sqrt{d/\delta}$ should be satisfied.
In the particular case $t_f = (2/\omega_0)\sqrt{d/\delta}$, the maximum velocity
is $v_0= \omega_0 \sqrt{d\delta}$, thus $t_1= t_f/2$ and $t_2=0$.
Like for displacement minimization, the trajectory in this case is
reduced again to that of the time-optimal control problem.
Moreover, to make $t_1$ non-negative, $t_f$ should be less than
$(\sqrt{6}/\omega_0)\sqrt{d/\delta}$. If $t_f > (\sqrt{6}/\omega_0)\sqrt{d/\delta}$,
the optimal trajectory is the one in the unbounded-control case, as commented before. In other words,
$\delta > \delta_0 = 6 d /\omega^2_0  t^2_f $.
As a result, the segmented form in Eq. (\ref{control function-energy}) applies for
the interval $4 d/\omega^2_0 t^2_f \leq \delta \leq 6 d/\omega^2_0 t^2_f$ marked by vertical lines in Fig. \ref{energy}.
There is no solution for smaller times,
whereas the solution becomes the one for unbounded control for larger times.


In this energy-optimal trajectory, the time-averaged potential energy $\overline{E_p}$ should be minimized. The cost function in Eq. (\ref{costfuntion-E}) becomes
\beq
J_E = m \omega^2_0 \delta^2 t_1 +\frac{1}{6} m \omega^2_0 \delta^2 t_2,
\eeq
and therefore
\beq
\overline{E_p}= \frac{\int^{t_f}_0 E_p dt}{t_f}= m \omega^2_0 \delta^2 \left( \frac{2 t_1}{3 t_f}  +\frac{1}{6}\right),
\eeq
which finally results in
\beq
\label{energy-e}
\overline{E_p}=  \frac{1}{2} m \omega^2_0 \delta^2 \left( 1 - \frac{2\sqrt{3}}{3} \sqrt{1 -  \frac{4d}{ \omega^2_0  t^2_f \delta}} \right).
\eeq

In Fig. \ref{energy}, we compare this to the (larger) average energy
for the displacement-optimal problem, Eq. (\ref{energy-d}),
and the lower bound Eq. (\ref{energy bound}), and also demonstrate that
the lower energy bound can be realized when $t_f > (\sqrt{6}/\omega_0)\sqrt{d/\delta}$.
%
%
%
%
%
%
%
%
%
\section{Discussions and Conclusions}
We have proposed optimal protocols for fast atomic transport in harmonic traps combining the invariant-based inverse engineering method and optimal control theory. Optimal trajectories with ``bang-bang" and ``bang-off-bang" forms are respectively obtained for time-minimization and displacement-minimization with constrained displacement between the trap center and the center of mass of the particle density.
The transient energies for bounded and unbounded displacement are also minimized.

In the time-optimal problem, the minimal time, Eq. (\ref{tf-bangbang}), corresponds to a fixed constraint $\delta$.
Consistently with this, no solutions are found for displacement and energy minimization problems
for transport times shorter than the minimal time, i.e. for $t_f < (2/\omega_0)\sqrt{d/\delta}$.
To achieve fast and faithful transport in shorter times, an ``energy price'' must be paid by increasing $\delta$ which, in real traps, will also produce errors because of anharmonicities. The relation between the minimal (time-averaged) energy and the transport time $t_f$
obtained here is not at all trivial, in particular they are not simply inversely proportional,
see e.g. Eqs. (\ref{energy bound}) or (\ref{energy-e}),
as one might naively expect from the form of time-energy uncertainty relations.
The scaling laws found are also peculiar of transport. For example the minimal energy in Eq. (\ref{energy bound}) depends on $t^{-4}_f$ instead of the
$t^{-2}_f$ dependence applicable to engineered trap expansions \cite{energy}.

In a previous work on invariants and transport \cite{transport}, the energy bound for $\overline{E_p}$ was found using the Euler-Lagrange equation.
Here we have shown how to realize this bound by allowing the discontinuous acceleration of the trap at $t=0$ and $t=t_f$ in the unbounded control optimization.
In principle these and other discontinuities found
could be avoided by imposing appropriate bounds and using a powerful pseudospectral numerical optimization method \cite{stef,Li-JCP,Li-PNAS} to address the corresponding more complex optimal control problem.

Anharmonicity could be dealt with in a completely different way using the protocols
for anharmonic transport described in \cite{transport}, which require a compensation of inertial forces in the frame of the trap. This may be feasible or not depending on the accelerations
imparted and the corresponding optimization will be considered elsewhere.

Last but not least, the present results may be extended to Bose-Einstein condensates following \cite{transport2}. Tonks-Girardeau gases could also be treated with a simple generalization \cite{Chen}.

\section*{Acknowledgments}
We acknowledge funding by the Basque Government
(Grant No. IT472-10) and Ministerio de
Ciencia e Innovaci\'on (FIS2009-12773-C02-01).
X. C.  acknowledges financial support from Juan de la Cierva Programme and the National Natural Science Foundation of China (Grant No. 60806041);
E. T. from the Basque Government (Grant No. BFI08.151).
J.-S. Li thanks the AFOSR Grant FA9550-10-1-0146 for supporting this work.

\end{document}